\documentclass[aps,prl,amsmath,amssymb,preprint,%
]{revtex4-1}

\usepackage{graphicx}% Include figure files
\usepackage{dcolumn}% Align table columns on decimal point
\usepackage{bm}% bold math
\usepackage{xcolor}
\usepackage[utf8]{inputenc}
\usepackage[T1]{fontenc}
\usepackage{mathptmx}

\begin{document}

\preprint{AIP/123-QED}

\title[Optical Rectification and Thermal Currents in Optical Tunneling Gap Antenna]{Optical Rectification and Thermal Currents in Optical Tunneling Gap Antennas}
% Force line breaks with \\

\author{M. M. Mennementeuil}
 \affiliation{Laboratoire Interdisciplinaire Carnot de Bourgogne CNRS UMR 6303, Université de Bourgogne Franche-Comté, 21000 Dijon, France}%Lines break automatically or can be forced with \\
 \author{M. Buret}
 \affiliation{Laboratoire Interdisciplinaire Carnot de Bourgogne CNRS UMR 6303, Université de Bourgogne Franche-Comté, 21000 Dijon, France}
 
\author{G. Colas des Francs}
\affiliation{Laboratoire Interdisciplinaire Carnot de Bourgogne CNRS UMR 6303, Université de Bourgogne Franche-Comté, 21000 Dijon, France}
\author{A. Bouhelier}
  \email{alexandre.bouhelier@u-bourgogne.fr}
\affiliation{Laboratoire Interdisciplinaire Carnot de Bourgogne CNRS UMR 6303, Université de Bourgogne Franche-Comté, 21000 Dijon, France}

\date{\today}% It is always \today, today,

\begin{abstract}
Electrically-contacted optical gap antennas are nanoscale interface devices enabling the transduction between photons and electrons. This new generation of devices captures visible to near infrared electromagnetic radiation and converts the incident energy in a direct-current (DC) electrical signal. The nanoscale rectenna is usually constituted of metal elements (e.g. gold). Light absorption by the metal contacts may lead to additional thermal effects which need to be taken into account to understand the complete photo-response of the device. The purpose of this communication is to discuss the contribution of laser-induced thermo-electric effects in the photo-assisted electronic transport. 
\end{abstract}

\maketitle

\section{\label{sec:introduction}Introduction}

Harvesting of electromagnetic energy has been fueling the quest for developing alternative technologies to compete with semiconductor-based photovoltaics. Amongst the different pretenders are metal-based rectifying antennas operating in the visible part of the spectrum~\cite{Cutler12,Moddel12}. These nano-devices combine in a footprint commensurate with the incoming wavelength the receiving antenna and the rectifying element in the form of a tunnel gap. Several advantages speak for such monolithic integration of the antenna and the rectifier. The electromagnetic response of the gap provides an enhanced interaction with the incident light~\cite{krenn03,natelson11prb}. The gap serves as capture antenna. The static electric polarity applied across the gap, geometrical and material asymmetries~\cite{Cuevas07,Mayer09}, as well as the ultrafast transit time across the tunneling barrier~\cite{Gabelli18} contribute to rectify high frequency radiation to an electrical DC signal. Here the gap acts as a rectifier for visible light~\cite{Gustafason75,Tucker79,Raynaud20}. This new family of devices found new application venues notably for on-chip detection of light signal~\cite{Hobbs2007,Nijhuis17,Dasgupta:18}. Optical rectification occurring at the level of an individual nano-scale rectenna has been observed repeatedly by a direct illumination of the tunneling gap~\cite{Selzer09,ward2010,selzerNL11,Stolz2014}.

However, rectification of the incident light may not be the only mechanism providing a photo-generated electrical signal. A portion of the light is inevitably absorbed by the metal parts, which lead to an increase of the system's temperature. Expansion of the electrodes, thermo-induced voltages and currents are therefore likely to affect the conductance of the tunneling gap and contribute to a concurrent current flow summed with the rectified response. While combining different mechanisms for generating a current may be of advantage to increase the overall efficiency of the device, the temporal dynamics of thermal effects are many orders of magnitude slower than optical rectification and are thus limiting the attainable response bandwidth. These adverse thermal effects were preponderant in the context of photo-assisted scanning tunneling microscopy~\cite{Grafstrom02}, but were generally discarded for planar devices on account of a better heat dissipation via the substrate~\cite{Selzer09,ward2010}. However, reports also demonstrated the occurrence of photo-thermal electrical responses in nanoscale junctions and constrictions~\cite{mceuen11,kopp12,Stolz2014,Sheer16,Natelson_JPCL17,Mennemanteuil18} as well as a sensitivity of thermo-electrical power to atomic gap configurations~\cite{Taniguchi_SR13}, and biasing polarity~\cite{Lee2013,Mennemanteuil18}. The purpose of this communication is to analyse the origin of the electrical signals generated by a laser-illuminated optical rectenna and to provide general guidelines to mitigate thermal contributions. 

\section{\label{sec:rectenna}Fabrication and characterization of tunneling gap optical antennas}

\subsection{\label{sec:fabrication}Fabrication procedure}
At the heart of the optical rectennas discussed here is a sub-nanometer tunneling barrier separating two metal electrodes. For optical characterization purposes and ease of fabrication, we fabricate in-plane tunnel gaps by conducting a controlled electromigration~\cite{Kim2021} of two tapered Au electrodes bridged by a nano-constriction. These parts are realized by electron-beam lithography on a glass coverslip and subsequent thermal evaporation of a 3~nm thick Cr adhesion layer and a 60~nm thick Au layer. The typical dimensions of the constriction is 200~nm$\times$200~nm. Macroscopic contact electrodes used to electrically interrogate the device are then fabricated by standard optical lithography followed by the same metal evaporation processes. To electromigrate the constriction, we proceed with a methodology described elsewhere~\cite{Dasgupta:18b}. In brief, we apply a voltage $V$ across the two contact electrodes and monitor the time evolution of the constriction's dynamic conductance. The onset of electromigration is characterized by a rapid decay of the conductance $G$, and generally occurs for an applied voltage of about $V=2$~V. To contain the catastrophic runaway of the process, $V$ is immediately reduced. The amplitude of the voltage is then either slightly increased to trigger again the atomic rearrangement and thinning down of the constriction or left constant if Joule heating imparted by the current flowing in the device is sufficient to assist the electromigration. This process is cycled until the conductance reaches the tunneling regime characterized by $G<G_{\rm 0}$, where $G_{\rm 0}=2e^2/h=77~\mu$S is the quantum of conductance, $e$ is the electron charge and $h$ is Planck's constant.  

\begin{figure}
\includegraphics{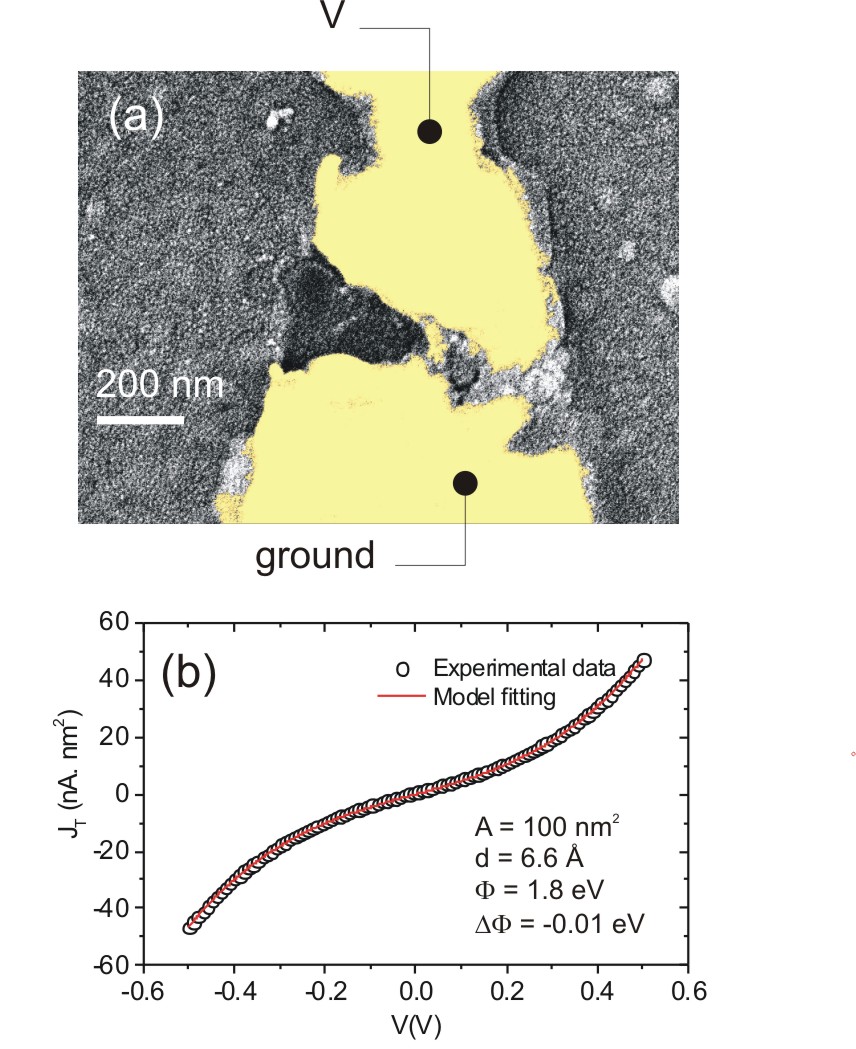}
\caption{\label{fig:SEM} (a) Scanning electron micrograph of a junction realized by electromigration. Au is represented by a yellow hue. (b) Typical output characteristic and a fit to the data. The extracted parameters are the gap distance $d$, the average barrier energy $\Phi$, and its asymmetry $\Delta \Phi$.}
\end{figure}

A typical scanning electron micrograph of a junction is displayed in the false-color image of Fig.~\ref{fig:SEM}. The contours of the broken section are ill-defined because of the stochastic nature of the electromigration. We generally observe that the rupture of the gold electrode does not necessarily occurs at the constriction, but is slightly displaced toward the source electrode~\cite{Dasgupta:18b}. Note that for the purpose of electron imaging, a thin Au conductive layer has been deposited on the entire sample after being tested.   

\subsection{\label{sec:electrical characterization}Electrical characterization}

To have a complete picture of the rectenna's operation, we electrically interrogate the output characteristic of the tunneling barrier. We treat the barrier in the framework of Simmons' model of transport~\cite{simmons63} and apply the methodology described by Dasgupta et al.~\cite{Dasgupta:18b}. In the limit of a small electron kinetic energy $eV$ compared to the average barrier height $\Phi$, the conductance dependence with the voltage bias $V$ applied across the gap takes an analytical form:~\cite{brinkman70}

\begin{equation}
\begin{split}
G(V) = G(0) \left(1 - \frac{\Gamma \Delta \Phi}{16 \Phi ^{\frac{3}{2}}} V + \frac{9\Gamma^{2}}{128} V^{2}\right),
\end{split}
\label{simmons}
\end{equation}
where $\Delta \Phi=\Phi_1-\Phi_2$ represents the difference of the barrier height at the two sides of the gap, $\Gamma = 4\sqrt{2m}d/3 \hslash$ with $m$ the electron mass and $d$ the gap distance. $G(0)$ writes:~\cite{brinkman70} 

\begin{equation}
\begin{split}
G(0) = A \times 3.16 \cdot 10^{-4} \frac{\sqrt{\Phi}}{d} \exp \left[-1.025 d\sqrt{\Phi}\right].
\end{split}
\label{G(0)}
\end{equation}
$A$ is the effective area of the junction in nm$^2$, $d$ is in \AA, $V$ is in V, and $\Phi$ is in eV.

The current density flowing through the barrier $J(V)$ is thus
\begin{widetext}
\begin{equation}
\begin{split}
J(V)&= \int_{0}^{V} G(V) dV\\
&= \left( \frac{3.16\cdot 10^{-4} \sqrt{\Phi}}{d} V - \frac{2.7\cdot10^{-5} \Delta \Phi}{\Phi} V^{2} + \frac{5.5\cdot10^{-5} d \sqrt{\Phi}}{\Phi} V^{3}\right) \exp \left[{-1.025d \sqrt{\Phi}}\right].
\end{split}
\label{eq:Simmon}
\end{equation}
\end{widetext}

The set of parameters [$d, \Phi, \Delta \Phi$] can be extracted by fitting the experimental output characteristic $I(V)$ with Eq.~\ref{eq:Simmon} and fixing the junction's area $A$. Figure~\ref{fig:SEM}(b) shows a typical output characteristic of an electromigrated junction together with a fit of the experimental data points using Eq.~\ref{eq:Simmon} and setting $A=10\times10$~nm$^2$. In this example, the estimated gap is 6.6~\AA~ and the barrier is weakly asymmetric with respect to the applied polarity (negligible $\Delta \Phi$). The extracted barrier height ($\Phi=1.8$~eV) is much smaller than the work function of Au, but is consistent with previous reports of reduced barrier height in similar systems~\cite{mangin:09b,Hecht15,Stolz2014, Frimmer14, Dasgupta:18b}. 

\subsection{\label{sec:optical}Laser-induced signals}
To interrogate the response of the device upon laser illumination, the electromigrated junction is placed on an inverted microscope equipped with a high numerical aperture (NA) oil immersion objective lens (NA=1.49, 100$\times$). A 785~nm constant-wave (CW) laser is focused to a diffraction-limited area by the objective.  Then sample is raster scanned through the focal region to reconstruct maps of the different optical and electrical responses of the rectenna. We use a transimpedance amplifier to measure the current at the output of the device. The total current $I$ is the sum of the bias-induced tunnel current $I_{\rm b}$ and the photo-generated current $I_{\rm phot}$: 

\begin{equation}
I=I_{\rm b}+I_{\rm phot}
\end{equation}

\begin{figure*}
\includegraphics[width=18cm]{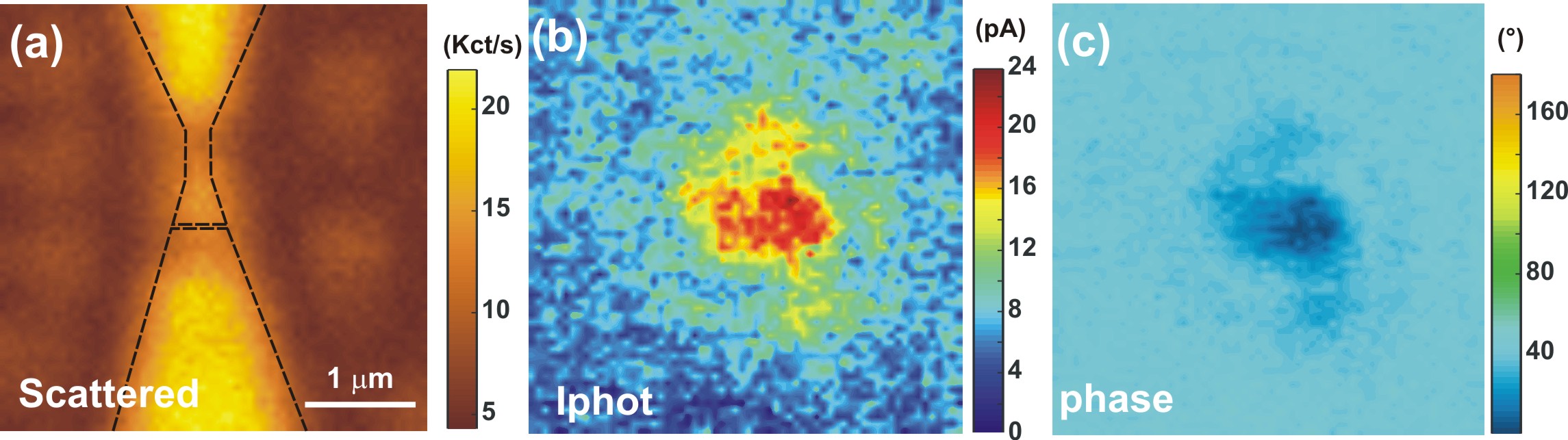}
\caption{\label{fig:confocal} (a) Confocal map of the back-scattered laser intensity. The outline of the tapered electrode, the constriction and the electromigrated region are marked by the dotted lines. (b) and (c) Simultaneously measured modulus $\lvert I_{\rm phot}\rvert$ and phase $\zeta$ of the laser-induced current. No electrical bias is applied here: $V=0$~V.}
\end{figure*}

At $V=0$, $I_{\rm b}$ is null and only $I_{\rm phot}$ exists. However, if the junction is externally biased, the contributors of the right hand side of the above expression are added. To extract $I_{\rm phot}$ from the total current regardless of the electrical biasing condition, we introduce a lock-in detection by chopping the laser beam at frequency of 831 Hz. The chopper provides a reference signal to sync the lock-in amplifier. 

An example of the junction's responses is illustrated in the confocal maps of Fig~\ref{fig:confocal}. Figure~\ref{fig:confocal}(a) is a map of the laser intensity partially back-reflected from the sample as it is scanned through the focus. The reflected laser intensity is recorded by an avalanche photo-diode (APD) placed in a conjugate object plane of the microscope. This reflection map helps us to identify the device geometry because Au surfaces give a higher reflected laser signal. The electrodes and the position of the tunnel junction are approximately outlined by the dotted lines. In this experiment the laser intensity at the focal spot is estimated at 486 kW$\cdot$cm$^2$ and a density filter attenuates the reflected beam detected by the APD. We simultaneously record the modulus and phase outputs of the lock-in amplifier, $\lvert I_{\rm phot}\rvert$ and $\zeta$. These two signals are displayed in Fig.~\ref{fig:confocal}(b) and (c) for $V=0$, respectively. When the electromigrated gap overlaps the laser focus, there is a photo-current of approximately 20 pA generated. The lateral extension of the response results from a convolution between the diffraction-limited area of the excitation spot and the capture cross-section of the responsive region, and perhaps a residual over or under focus position of the objective lens. The phase signal displayed in Fig.~\ref{fig:confocal}(c) stays approximately in-phase ($\zeta=0$) whenever a photo-signal is detected. This indicates that the direction of the current remains constant regardless of the laser position. The question at this stage of the discussion is to identify the possible processes contributing to $\lvert I_{\rm phot}\rvert$ when the laser overlaps the feedgap region of the device.

\section{\label{sec:contribution} Contributing laser-induced processes}

In the following sections we review the different mechanisms that could potentially contribute for generating a laser-induced current in the tunnel junction.

\subsection{Optical rectification, from a classical perspective}
\begin{figure}
\includegraphics[width=8cm]{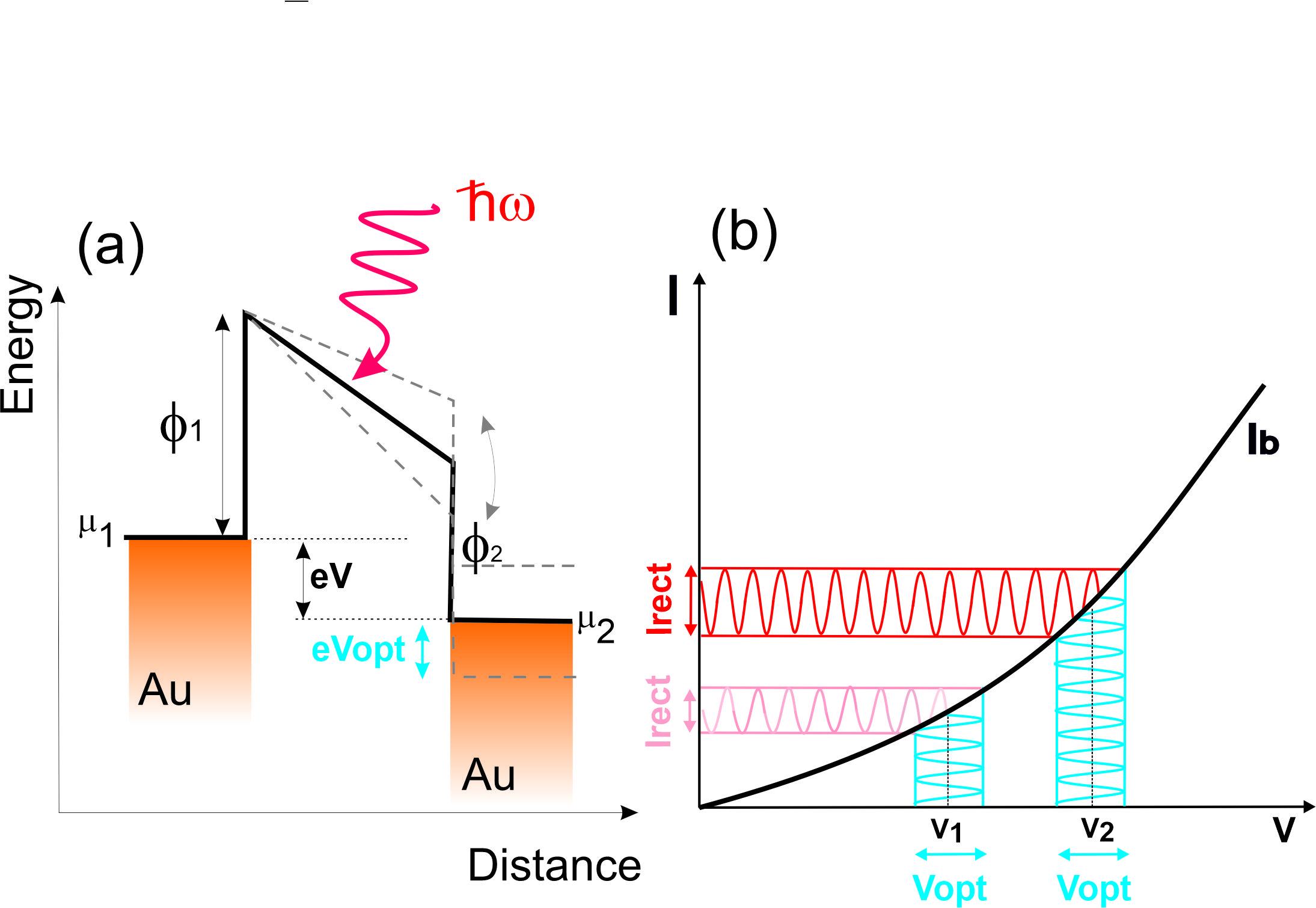}
\caption{\label{fig:rectification} (a) Energy diagram picturing a tunnel barrier biased at illuminated by an electromagnetic field with a photon energy $\hbar\omega$. The oscillation of the barrier at $\omega$ creates an optically-induced voltage $V_{\rm opt}$. $\mu_{1}$ and $\mu_{2}$ are the chemical potential of the two Au electrodes and $\phi_{1}$ and $\phi_{2}$ are the effective work functions. (b) Sketched output characteristics showing the amplitude of the rectified current $I_{\rm rect}$($V_{\rm opt}$) for two biasing operating set-points $V_{1}$ and $V_{2}$.}
\end{figure}

Rectification of the electromagnetic field can be cast from classical concepts ruling the operating mode of microwave rectennas. An illustration of the energy landscape of a generic biased rectenna is depicted in Fig.~\ref{fig:rectification}(a). A potential barrier between two metallic electrodes whose electrochemical potentials $\mu_{1}$ and $\mu_{2}$ are separated by an energy $eV$, undergoes additional oscillating potential $V_{\rm opt}$ in the presence of an incident radiation at an energy $\hbar \omega$. Figure~\ref{fig:rectification}(b) illustrates the effect of the working point $V$ of the device output's characteristics on the amplitude of the rectified current. The non-linear evolution of the $I_{\rm b}$(V) curve provides an handle to control the magnitude $I_{\rm rect}$ with the external static bias $V$ for the same optical voltage $V_{\rm opt}$ produced at the gap.

Following a classical description of the rectification and neglecting for now any additional thermal contributions, the total bias at the terminal of the rectenna is
\begin{equation}
V_{\rm rect} (t) = V + V_{\rm opt} \cos(\omega t)
\end{equation}

The current $I_{\rm rect}(t)$ induced by the optical potential $V_{\rm opt}$ can then be obtained by a Taylor expansion: \cite{Cutler12}

\begin{equation}
\begin{split}
I_{\rm rect}(t) &= \sum_{n=0}^{+\infty} \frac{1}{n!} \frac{d^{n} I_{\rm b}}{dV^{n}} [V_{\rm rect}(t) - V]^{n} = \sum_{n=0}^{+\infty} \frac{1}{n!} \frac{d^{n} I_{\rm b}}{dV^{n}} [V_{\rm opt}\cos(\omega t)]^{n}
\end{split}
\label{Eq:Irect}
\end{equation}

Using the trigonometric relation $V_{\rm opt}^{2} \cos^{2}\omega t = \frac{V_{\rm opt}}{2} [1 + \cos(2\omega t)]$ and similar relations developed at higher orders, the tunnel current $I_{\rm rect}(t)$ is described by the relation :

\begin{equation}
I_{\rm rect}(t)= \sum_{n=0}^{+\infty} I_{n} \cos n \omega t
\end{equation}

Considering only the term corresponding to the lowest order, the time averaged tunnel current through the nanojunction under light excitation and voltage $V$ is written :

\begin{equation}
I = I_{\rm b} + \frac{1}{4} V_{\rm opt}^{2} \frac{d^{2}I_{\rm b}}{dV^{2}} = I_{\rm b} + I_{\rm rect}
\label{eq:I_average}
\end{equation}

The first term corresponds to the current produced by the voltage $V$ applied to the rectenna while the second term corresponds to the additional current rectified by the device. This expression was used by Tu \textit{et al.}~\cite{Tu06} to analyse their experimental data on rectification at microwave frequencies. Equation~\ref{eq:I_average} was also used by Bragas~\cite{Bragas98} and then Natelson~\cite{ward2010} to identify and distinguish the contribution of the rectified current from the total current for rectenna illuminated in the visible spectral region.

Expression~\ref{Eq:Irect} assumes that the current flowing through the rectenna follows instantaneously the applied potential. However, this assumption is no longer valid when the period of the oscillations is comparable to the transit time of the electrons in the nanojunction, which is typically the case at optical frequencies~\cite{Mayer2011}. The classical description of the rectification process is therefore only applicable for devices operating at low frequencies or when the optical voltage $V_{\rm opt}$ created by the barrier oscillation is much lower than $\hbar\omega/e$. Formally, the interaction of the rectenna with an optical field  requires a quantum or semi-classical interpretation based on the photo-assisted tunneling (PAT)~\cite{Tien,Tucker79,Moddel12}. This description is provided in the appendix for completeness.  We will see in the following of this study that electromigrated junctions have a sufficiently low non-linearity with respect to $\hbar\omega/e$ to consider the classical model even in the presence of a high frequency excitation.

\subsection{Thermal effects}

When the laser illuminates the rectenna, and by extension its electrical feed-through, one should take into account the accompanying thermal effects promoted by the absorption of light. For sub-nanometer junctions, thermal expansion of the leads is likely to affect the magnitude of the tunneling current by closing the gap separating the two electrodes. This mechanism is particularly adverse in scanning tunneling microscopy~\cite{Grafstrom02} but was found to be negligible in device anchored to a substrate and illuminated with laser energy much lower than the material's interband transitions~\cite{Selzer09}. Another contributor is induced by temperature gradient that may exist between the two sides of the tunneling junction. A difference in temperature $\Delta T$ introduces an asymmetric electronic distribution, which establishes a thermo-current across the gap under closed-circuit condition. Even for devices constituted of homogeneous material, thermo-electric currents and voltages were systematically observed in similar device configurations~\cite{Taniguchi_SR13,Natelson_JPCL17,Mennemanteuil18,Natelson_ACS20}. By taking these additional laser-induced thermal contributions into consideration, the photo-induced current $I_{\rm phot}$ may be written as a sum of several contributions depending on the laser intensity $P_{\rm laser}$ including one corresponding to the rectification of the optical field $I_{\rm rect}$, and others imparted by thermal effects such as electrode expansion, $I_{\rm \delta d}$, and thermo-current $I_{\rm TE}$.  A variation of the optical potential $V_{\rm opt}$ and of the thermo-voltage $V_{\rm TE}$ will lead to a change of the total current dictated by the output characteristics $I(V)$. The total current of an illuminated biased junction is thus
\begin{equation}
\begin{split}
I &=I_{\rm b}+I_{\rm phot}\\
&=I_{\rm b}(V, d)+I_{\rm rect}(V, V_{\rm opt},d) +I_{\rm \delta d}(P_{\rm laser},d)+I_{\rm TE}(V, \Delta T,d).
\end{split}
\label{eq:current_total}
\end{equation}
For a given applied bias $V$, Eq.~\ref{eq:current_total} clearly shows that a modification of the junction size $d$ by the laser would affect all the contributing terms. This is therefore a sensitive contribution that needs to be taken into account. Concerning the thermo-current, we can make the following assumption: 
for a laser focused and centered on the junction, the temperature raise on both sides can be considered symmetric as the size of the diffraction-limited focal region (ca. 260 nm) is overlapping a significant portion of the electrodes. Local structural differences at the vicinity of the electromigrated junction (Fig.~\ref{fig:SEM}(a)) are probably irrelevant. The confocal image in Fig.~\ref{fig:confocal}(a) showing the back-scattered laser signal does not reveal any local responses specific to the electrode geometry. Under this particular excitation condition (centered, symmetric, equal temperature on both side of the gap), the literature consistently shows a vanishing thermopower~\cite{Natelson_JPCL17,Mennemanteuil18} for unbiased devices.

\section{\label{sec:experiments} Estimating the different contributions}

In the following sections, we investigate the evolution of the different signals as a function of laser intensity and applied bias in order to identify the processes at play contributing to the measured currents. 

\subsection{Evolution of the tunnel current with laser intensity}

\begin{figure}
\includegraphics[width=8cm]{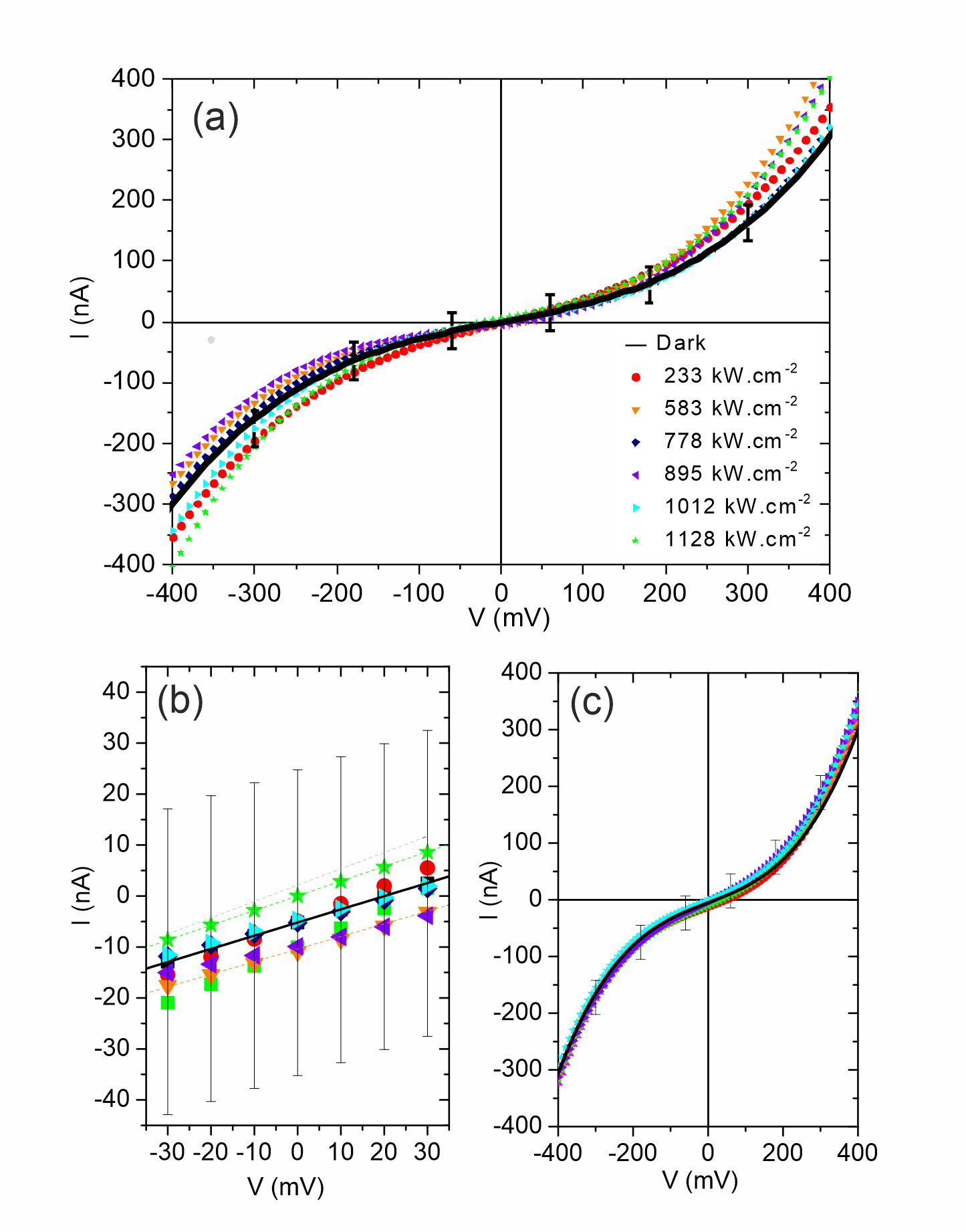}
\caption{\label{fig:Ivslaser} (a) $I(V)$ characteristic of the junction under various laser intensity. (b) Zoom in of the curve around $\pm$ 30~mV where the evolution of current evolves linearly with applied voltage. The dashed lines are linear fits to the data taken at different excitation intensities used to estimate the static conductance. (c) $I(V)$ characteristics when the laser is positioned on one of the electrode, away from the junction.}
\end{figure}

In this section we interrogate how the total current contributing to Eq.~\ref{eq:current_total} flowing through the junction is affected by the laser light. The output characteristic $I(V)$ of the electromigrated junction measured under laser illumination is shown in Fig.~\ref{fig:Ivslaser}(a). The black solid line shows the output characteristic of the rectenna without any optical stimulus (dark response). When illuminated, the deviation of $I(V)$ curve becomes increasingly pronounced with laser intensity. This evolution can be placed in regard of the schematic depicted in Fig.~\ref{fig:rectification}(b) where the magnitude of the rectified current increases with the nonlinearity of the $I(V)$ curve. In the linear portion of the curve taken around the zero-bias condition (Fig.~\ref{fig:Ivslaser}(b)), the level of the noise prevents us to establish a consistent trend with laser intensity in these measurements. 

The evolution of the conductance and the nonlinearity of the output characteristic is a marker to assess the stability of the tunnel junction during the measurement. Increasing $V$ or $P_{\rm laser}$ above the damage threshold will inevitably lead to a decrease of the conductance by a widening of the gap distance $d$ and thereby changing the nonlinearity of the device. A first confirmation of the stability of the junction is demonstrated in Fig.~\ref{fig:Ivslaser}(b). The linear fits (dashed curves) to the data indicates that the static conductance stays constant with laser intensity. A second demonstration is provided by the $I(V)$ curves displayed in Fig.~\ref{fig:Ivslaser}(c) for various laser power. Here, the laser is placed on the upper electrode (Fig.~\ref{fig:confocal}). There is no noticeable variation of the junction's electrical characteristics upon the different voltage sweeps and laser powers sampled suggesting that the laser, and the voltage, do not affect the geometry of the device.  This is further confirmed by the following measurements. 

According to Simmon's model of the transport (Eq.~\ref{eq:Simmon}), the exponential dependence of the tunneling current $I$ does not depends on applied voltage $V$ but is only driven by the size of the barrier $d$, and the effective work function $\Phi$. This exponential dependence is conserved in the first derivative and subsequent derivatives of $I(V)$. For the first and second derivative, the expressions read: 

\begin{equation}
\begin{split}
\frac{dI}{dV} = & A\left(\frac{3.1\cdot 10^{-4}\sqrt{\Phi}}{d}-\frac{-5.4\cdot 10^{-5}\Delta\Phi}{\Phi}V+\frac{1.65\cdot10^{-4}d\sqrt{\Phi}}{\Phi}V^2 \right) e^{-1.025d\sqrt{\Phi}}  \\
\frac{d^2I}{dV^2} = & A\left(\frac{-5.4\cdot 10^{-5}\Delta \Phi}{\Phi}+\frac{3.3\cdot 10^{-4}d \sqrt{\Phi}}{\Phi}V\right)e^{-1.025d\sqrt{\Phi}}
\end{split}
 \label{eq:derivative}
\end{equation}

Experimentally, we measure the first and second derivative of the $I(V)$ by a lock-in detection of the total current at the $n^{\rm th}$ harmonic of the sync frequency of a 14~mV sinusoidal voltage $V_{\rm AC}$ applied to the rectenna~\cite{ward2010,Stolz2014}. Demodulating a current at a frequency $n\cdot f$ is formally equivalent to a measure of the $n^{\rm th}$-derivative~\cite{Adler66}. Here, $f=12.6$~kHz. If $n=1,2$, the lock-in outputs $S^{\rm nf}$ provide a magnitude of the junction's dynamic conductance and nonlinearity and write:

\begin{equation}
\begin{split}
& S^{\rm 1f}=V_{\rm AC}\frac{dI}{dV}\\
& S^{\rm 2f}=\frac{1}{4}V_{\rm AC}^2 \frac{d^2I}{dV^2}
\label{eq:lockin}
\end{split}
\end{equation}

\begin{figure}
\includegraphics[width=8cm]{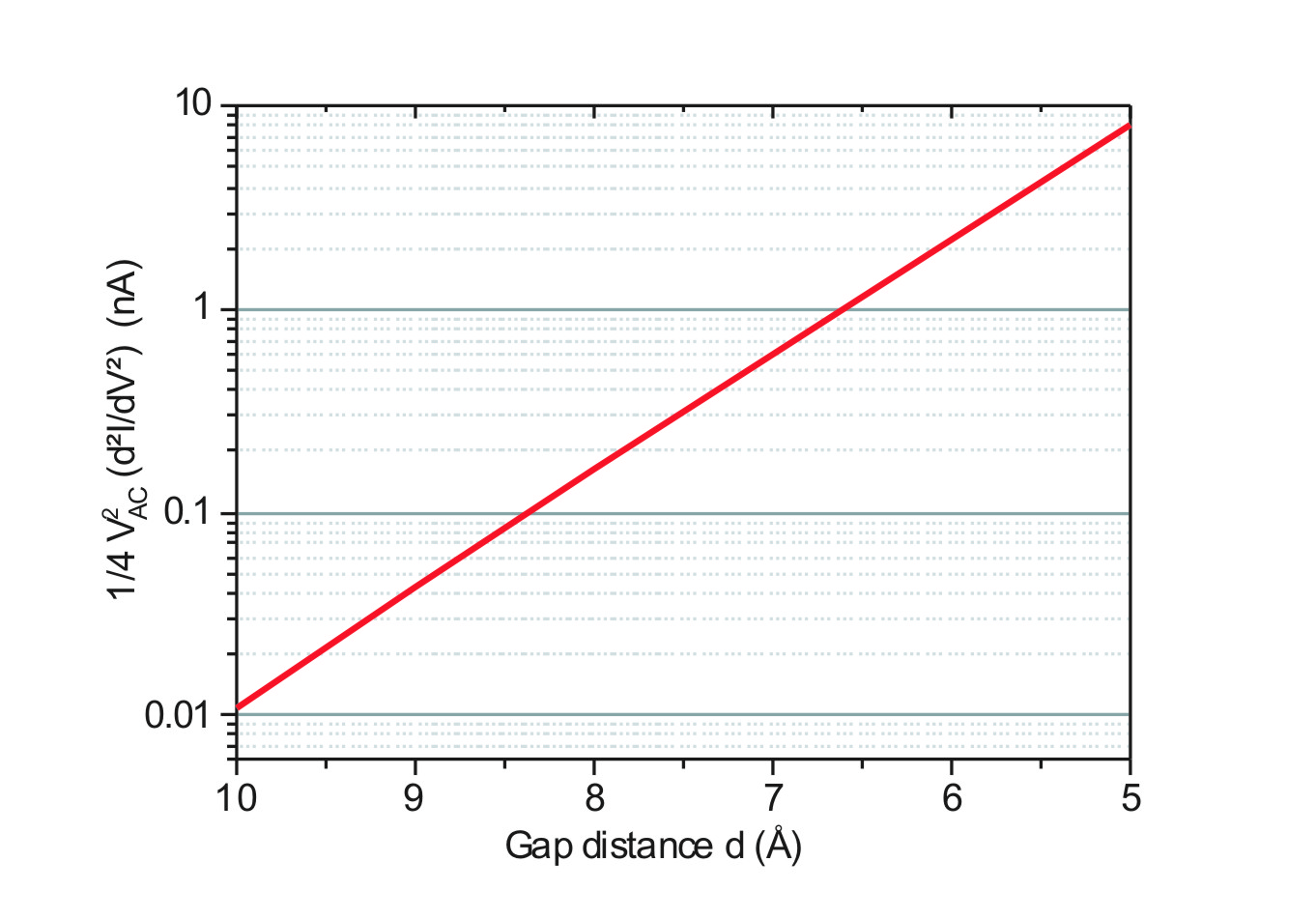}
\caption{\label{fig:d2Ivsd} (a) Semi logarithmic plot of the simulated evolution of the rectenna's nonlinearity probed by the output of a lock-in detection with varied gap size $d$.}
\end{figure}

 To appreciate how the nonlinearity of the $I(V)$ curve depends on small variations of the gap size $d$, we plot in Fig.~\ref{fig:d2Ivsd} the simulated lock-in output $S^{\rm 2f}$ as a function of $d$ for the junction parameters $A$, $\Phi$ and $\Delta \Phi$ deduced from the analysis of Fig.~\ref{fig:SEM}(b) and using Eq.~\ref{eq:derivative} and Eq.~\ref{eq:lockin}. We arbitrarily fix the applied voltage at $V=0.4$~V. Clearly, a small change of the gap size resulting from a thermal expansion or an atomic rearrangement due to the laser excitation or the operating voltage would introduce a large modification of the current' second derivative. 
 
\begin{figure}
\includegraphics[width=8cm]{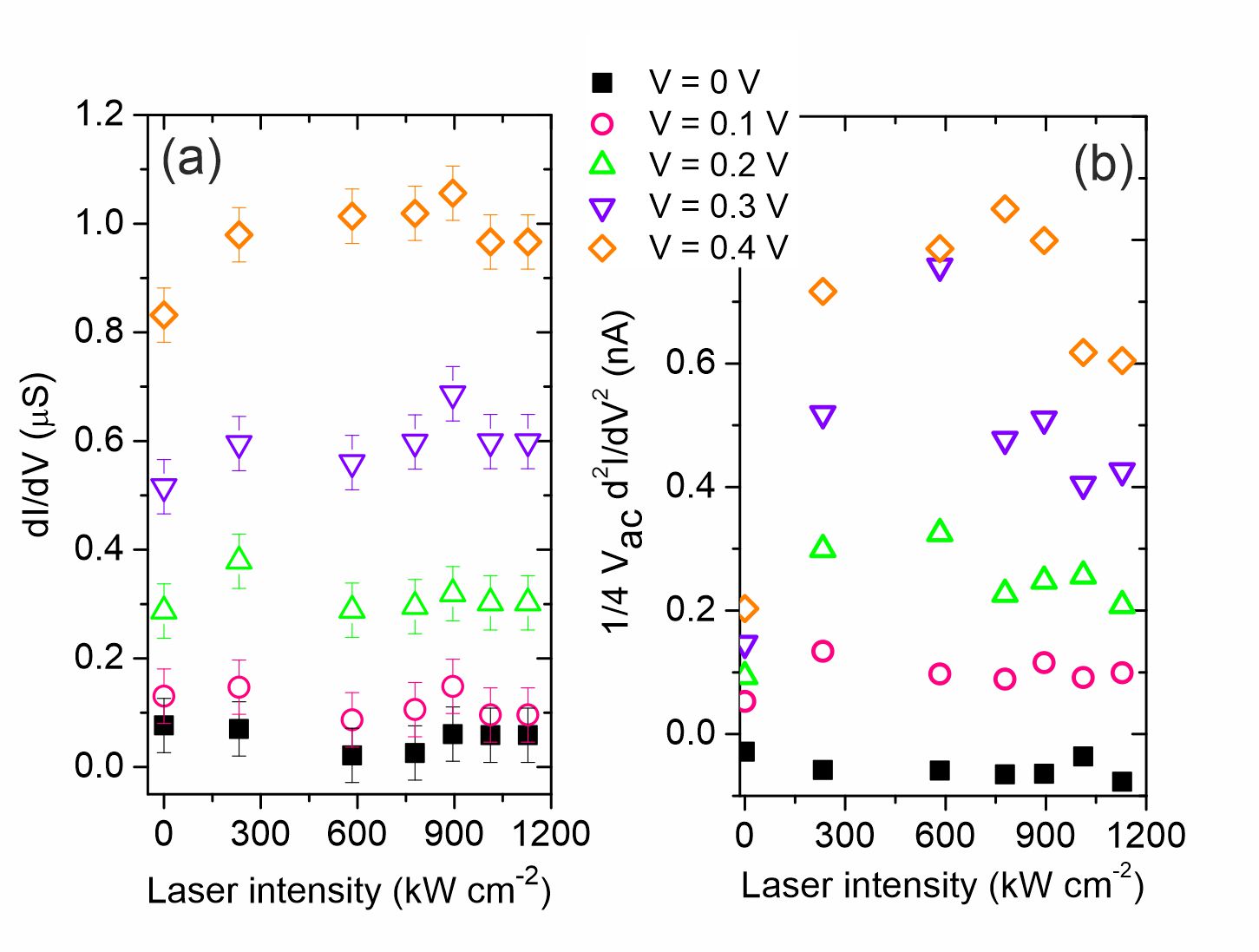}
\caption{\label{fig:Gvslaser} Evolution of the first (a) and the second derivatives (b) of the tunnel current measured by the lock-in with laser intensities for five bias set points $V$. (a) measures the dynamic conductance of the junction while (b) provides its nonlinearity.}
\end{figure}

Figure~\ref{fig:Gvslaser}(a) shows the dynamic conductance $G$ of the junction experimentally determined by normalizing the lock-in output $S^{\rm 1f}$ with $V_{\rm AC}$ for different laser intensities and operating voltages $V$. For the range of intensity interrogated, the dynamic conductance at null bias remains stable and confirms the static measurement performed by estimating the slopes of the $I(V)$ in Fig.~\ref{fig:Ivslaser}(b). When increasing the applied bias, the dynamic conductance increases because the device enters its nonlinear regime. Here too, the dynamic conductance does not show sensible variations with laser intensity indicating that the junction's intrinsic geometry is not altered by the illumination. 

We confirm the stability against closing of the gap size by measuring the second derivative of the output characteristics with laser intensity shown in Fig.~\ref{fig:Gvslaser}(b). For $V=0$~V, the junction's response is linear and the non-linearity is approximately null. When operating the device with voltage set-points greater than 100~mV, the characteristic cannot be considered linear and the different terms in $V^2$ and $V^3$ in Eq.~\ref{eq:Simmon} are becoming predominant. Concomitantly, the lock-in output $S^{\rm 2f}$ registers a larger signal indicative of a gaining non-linearity. The point is that this quantity is fairly steady within the range of laser intensity sampled. Compared to the trend expected if the laser would impart a linear expansion of the electrodes and a subsequent closing of the gap (Fig.~\ref{fig:d2Ivsd}), we conclude that the dependence of the tunnel current with laser intensity observed in Fig.~\ref{fig:Ivslaser} or the confocal response of Fig.~\ref{fig:confocal}(b) cannot be accounted for a modified gap size $d$ during illumination and voltage activation.

The conservation of the intrinsic electrical properties indicates that thermal expansion of the electrodes is likely to be negligible in our measurement and thus $I_{\rm \delta d}\approx 0$ in Eq.~\ref{eq:current_total}. The evolution of the total current with laser observed in the explored voltage range (Fig.~\ref{fig:Ivslaser}(a)) is then directly related to a current $I_{\rm phot}$ flowing through the rectenna and added to the tunnel current $I_{\rm b}$ imposed by the external potential.

\subsection{Evolution of the photo-current $I_{\rm phot}$ with laser intensity}

\begin{figure}
\includegraphics[width=8cm]{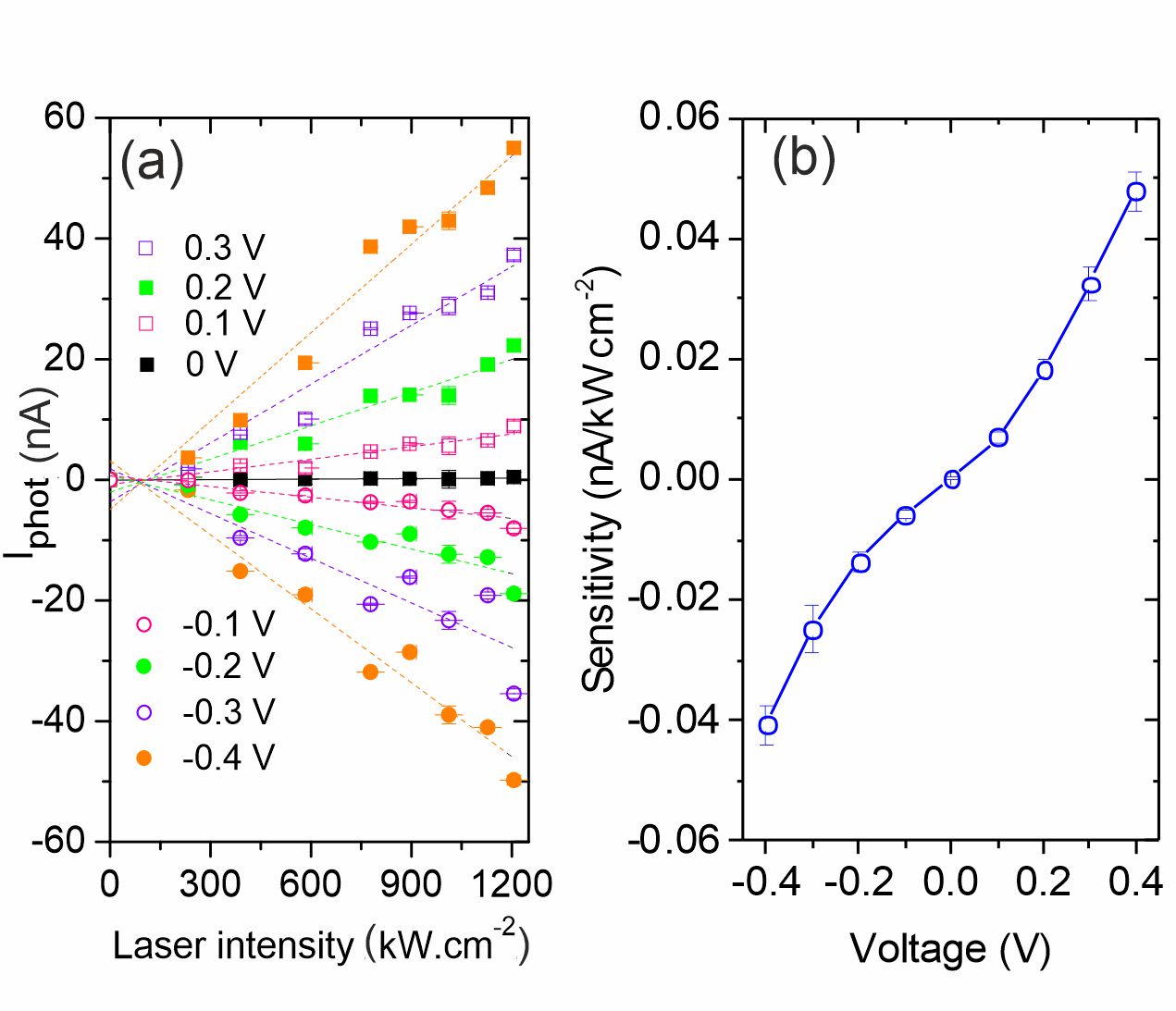}
\caption{\label{fig:Iphotvslaser} (a) Laser-induced current contribution $I_{\rm phot}$ delivered by the rectenna as a function of laser intensity for various symmetric voltage set-points. The dotted line are linear fits to the data. The slopes of the fits are reported in (b) for the corresponding voltages. The slope is a measured of the sensitivity of the device at a given bias set-point }
\end{figure}

In the following paragraph we specifically study the sensitivity of the laser-induced contribution $I_{\rm phot}$ of the total current to the laser intensity. For this purpose, we interrogate the lock-in output synchronized at the chopper frequency (see Sect.~\ref{sec:optical}) by changing the laser intensity and voltage set-points of the device. The results are displayed in Fig.~\ref{fig:Iphotvslaser}(a) and are represented by the data points. According to Eq.~\ref{eq:current_total}, $I_{\rm phot}$ is constituted of three different terms; one provided by the rectification of the optical field and the two others stemming from thermal effects.

We concluded the previous section by a negligible contribution $I_{\rm \delta d}$. $I_{\rm TE}$ is also neglected here because the symmetrical optical excitation of the rectenna and the even temperature on both sides of the gap. The remaining predominant term is thus $I_{\rm rect}$, which depends on the nonlinearity of the device and the square of the optical potential dropped across the gap $V_{\rm opt}^2$ (Eq.~\ref{eq:I_average}). In a first approximation $V_{\rm opt}$ is linked to the amplitude of the optical field via the relation $V_{\rm opt}=E_{\rm opt}\times d$. This simple relation may be weighted by local enhancement effect provided by the excitation of local surface plasmon resonances. The laser intensity $P_{\rm laser}$ generates the optical field following the relation $P_{\rm laser} =B \lvert E_{\rm opt} \rvert ^2$, where $B$ is a proportionality factor. $V_{\rm opt}^2$ is thus a linear function of the laser intensity with $V_{\rm opt}^2 \propto P_{\rm laser}d^2$. We can now understand the trend of $I_{\rm phot}$ featuring a linear dependency with $P_{\rm laser}$ in Fig.~\ref{fig:Iphotvslaser}(a) for all the operating voltages. The dotted lines are linear fits to the data. The fits are not constrained by a fixed intercept at $I_{\rm phot}=0$ because the data includes a dark noise contribution. The slope of the fits, which represents the sensitivity of the device, are reported in Fig.~\ref{fig:Iphotvslaser}(b) as a function of voltage $V$. As expected from Eq.~\ref{eq:I_average}, moving up the bias set-point increases the nonlinearity of the device and thus the term $d^2I/dV^2$ (Fig.~\ref{fig:Gvslaser}(b)) and by consequence the rectified signal $I_{\rm rect}$.

\begin{figure}
\includegraphics[width=8cm]{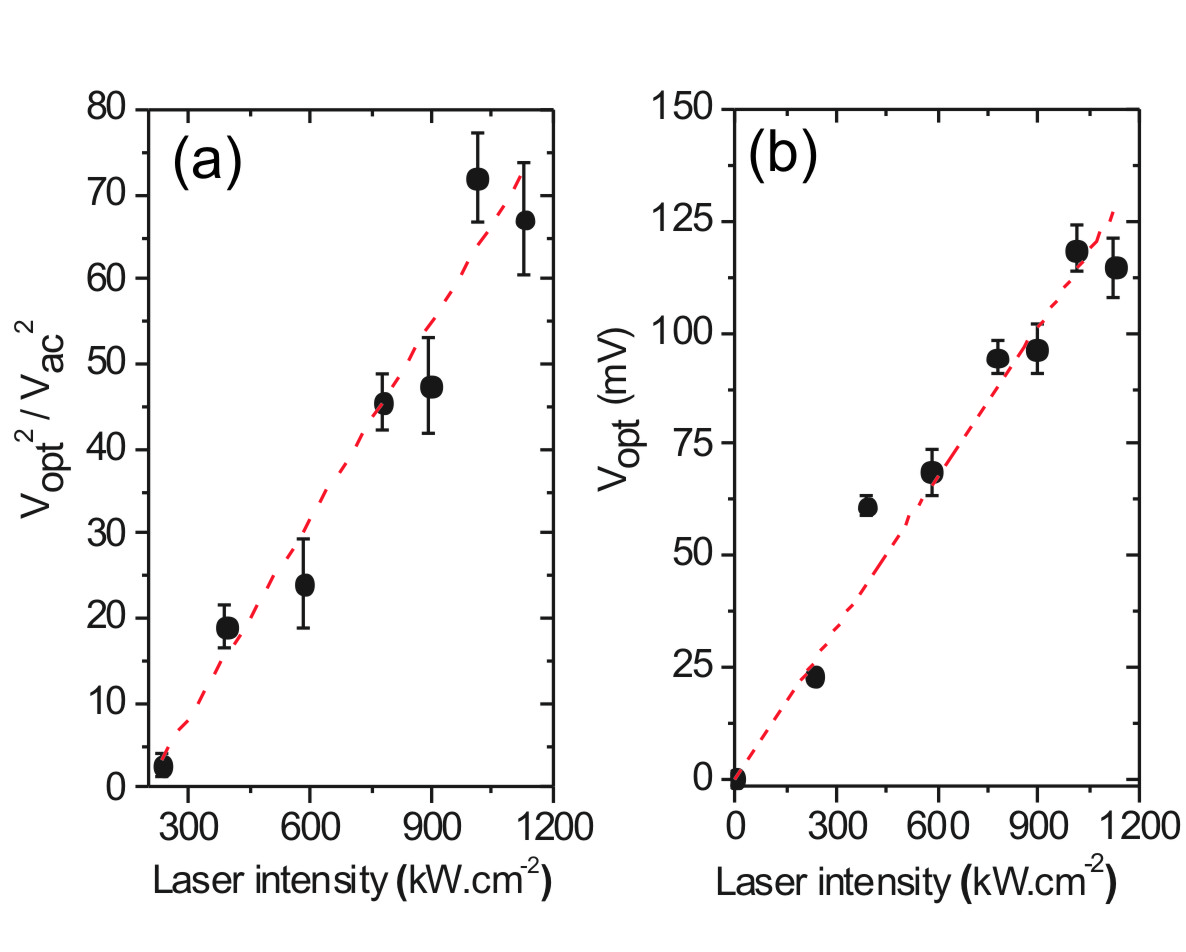}
\caption{\label{fig:vopt} (a) Ratio $V_{\rm opt}^2/V_{\rm AC}^2$ with laser intensity. (b) Deduced values of $V_{\rm opt}$ with $V_{\rm AC}=14$~mV. The dashed lines in both graphs are linear fits. The linear trends are consistent with the rectification picture.}
\end{figure}

Comparing the magnitude of the signals $I_{\rm phot}=I_{\rm rect}$ and $S^{2f}$ (Eq.~\ref{eq:lockin}) enables to estimate the amplitude of the optical $V_{\rm opt}$ generated by illuminating the gap with the laser. In Fig.~\ref{fig:vopt}(a), we make the ratio between $I_{\rm phot}/S^{2f} = V_{\rm opt}^2/V_{\rm AC}^2$. $ V_{\rm AC}$ is fixed experimentally at 14 mV. Here too, the ratio has a clear linear relationship with the laser intensity. Figure~\ref{fig:vopt}(b) shows the resulting optical potential $V_{\rm opt}$ as a function of the laser intensity. The inferred values are consistent with earlier an report of Ward \emph{et al.}~\cite{ward2010} and are within the validity of the semi-classical description of the rectification process since $eV_{\rm opt} \ll \hbar\omega$. In this study, the step-like evolution of the $I(V)$ upon the rectification process is not observed since the nonlinearity of the junction in the scale of $V_{\rm opt}$ is not sufficient to identify a quantized rectified current. The photo-assisted tunneling current can thus be described from a classical formalism.

\subsection{Example of a thermal effect}

\begin{figure}
\includegraphics[width=8cm]{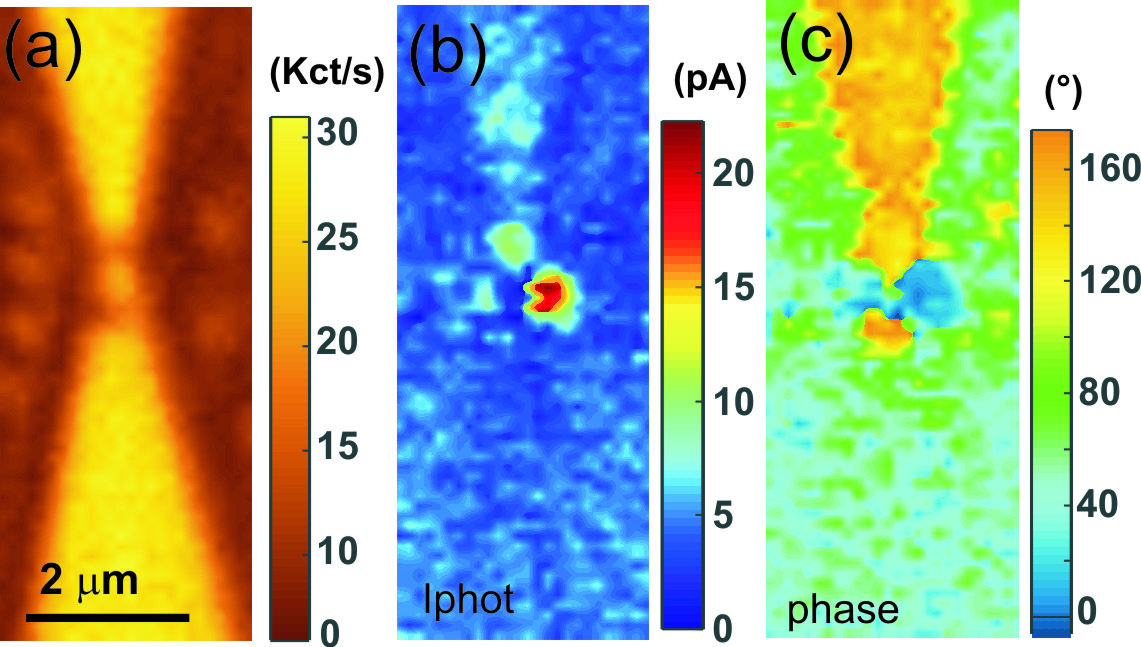}
\caption{\label{fig:confocalthermo} (a) Confocal map of the laser back-scattered intensity. The rounded features nearby the tapered sections are gold nanoparticles used in another study. (b) and (c) Simultaneously measured modulus $\lvert I_{\rm phot}\rvert$ and phase $\zeta$ of the laser-induced current. $V=0$~V and $P_{\rm laser}=1.09$~MW$\cdot$ cm$^2$.}
\end{figure}

In the previous sections, we argued that the thermal imbalance between the two metal electrodes can be neglected on account of a symmetric excitation, \emph{i.e} a laser focused on the rectifying feed. While this was true with the junction discussed above (see for instance the confocal maps of Fig.~\ref{fig:confocal}), the intricacy of the electromigration process and the resulting gap geometry sometimes lead to more complicated case with clear evidence of a current flow $I_{\rm TE}$ (Eq.~\ref{eq:current_total}) generated by a thermal gradient. Such an example is illustrated in the confocal maps of Fig.~\ref{fig:confocalthermo}. The outline of the device is pictured in the back-scattered laser intensity map of Fig.~\ref{fig:confocalthermo}(a) taken at a reduced laser power. With the diffraction-limited resolution of the microscope, there is no significant differences between this device and the previous junction's confocal response (Fig.~\ref{fig:confocal}(a)). The photo-current map of Fig.~\ref{fig:confocalthermo}(b) however shows a drastically different behavior. The laser of the power is here higher than in Fig.~\ref{fig:confocal} with $P_{\rm laser}=1089$~kW$\cdot$ cm$^{-2}$. Two distinct spots, one intense and a second one less pronounced, are now dictating the spatial response of the device. A weak photo-response is also detected when the laser impinges on the upper electrode, response not seen in the first case discussed in Fig.~\ref{fig:confocal}. This photo-current generated when the laser shines on an electrode is a signature of a thermo-current flowing in the device~\cite{Mennemanteuil18} created by laser-induced temperature difference. The phase map of Fig.~\ref{fig:confocalthermo}(c) confirms the thermal nature of the photo-current. There is a net and abrupt $\approx 170^\circ$ phase shift between the principal features in the $I_{\rm phot}$ map indicating a direction of the current which depends on the precise location of the laser with respect to the symmetry of the device. 

\section{\label{sec:conclusion} Conclusions}
To conclude, we reviewed and identified the different origins of the current photo-generated by planar optical rectennas. Because theses new family of devices consist of a tunneling junction formed between two metal electrodes, thermal effects inherent to light absorption must be considered. This is a key aspect to consider for deploying optical rectennas as ultrafast integrated opto-electronic units as the intrinsic dynamics behind optical rectification and thermal effects are differing many orders of magnitude. 

The differentiation of thermal processes from optical rectification phenomenon was carried out by electronic characterisation of rectifying feed under direct laser illumination. Tracking of the junction's dynamic conductance and the second derivative of its output characteristic enabled us to exclude any possibility of thermal expansion of the metal electrodes.  The rectenna intrinsic electrical characteristics is thus maintained throughout the experiment.

The extraction of photo-current $I_{\rm opt}$ from the total electrical current generated by the device under a symmetrical illumination was performed by a lock-in detection. We have shown that a linear evolution of the photo-current with the laser intensity indicates that the measured photo-current is a rectified contribution produced by the device. The ratio of the photo-current signal and the electrical nonlinearity of the rectenna allowed us to infer the optical voltage $V_{\rm opt}$ dropped at the rectenna's feed. We found values of a few tens of milliVolt. Considering these low $V_{\rm opt}$, the rectified current can be accurately described by a classical rectification formalism. The nonlinearity of the output characteristics at the scale of a few tens of millivolts is not sufficient to observe a quantization of the rectified current as expected by the semi-classical model usually used for high frequencies excitation.

We found that asymmetric illumination of the junction might cause the establishment of a thermal gradient. This is may affect the tunnel barrier electronic distribution due to a thermal imbalance between the two metal electrodes. An additional current is then generated and added on the rectified signal. The direction of this thermo-induced current flow can be deduced by extracting the phase of the measured photo-current. The current drifts according to the illuminated side of the junction.

\begin{acknowledgments}
The research leading to these results has received fundings from the European Research Council under the European Community’s Seventh Framework Program FP7/2007–2013 Grant Agreement 306772, the European Union through the PO FEDER-FSE Bourgogne 2014/2020 programs, the Conseil R\'egional de Bourgogne Franche-Comt\'e and has been supported by the EIPHI Graduate School (contract ANR-17-EURE-0002). Device fabrication was performed in the technological platform ARCEN Carnot with the support of the R\'egion de Bourgogne Franche-Comt\'e, the D\'el\'egation R\'egionale \`a la Recherche et \`a la Technologie (DRRT), and the CNRS. The data that support the findings of this study are available from the corresponding author upon reasonable request.
\end{acknowledgments}

\appendix

\section{Semi-classical description of the rectified current}

Photo-assisted tunneling of electrons considers changes in the density of electronic states of the potential barrier in the presence of an oscillating field. The optical field is then treated as a time-dependent perturbation. 
The perturbation $V_{\rm opt}$ is described through a total time-dependent Hamiltonian term $H$ such that :

\begin{equation}
H(t) = H_{0}(t) + eV_{\rm opt} \cos(\omega t)
\end{equation}

where $H_{0}$ is the Hamiltonian term characterising the tunnel barrier in absence of the AC disturbance and whose associated wave function is of the form : 
 
\begin{equation}
\Psi_{0} (x,y,z,t) = \psi(x,y,z) \exp {-iEt\slash\hbar}
\end{equation}

The AC disturbance induced by the optical radiation then introduces an additional phase term whose effect can be modeled by a new wave function :

\begin{equation}
\Psi (x,y,z,t) = \psi(x,y,z)\exp(-iEt\slash\hbar) \exp\left[ -(i\slash\hbar)\int^{t} dt'eV_{\rm opt} \cos(\omega t')\right]
\end{equation}

Time integrating and using the Jacobi-Anger expansion, the wave function becomes :

\begin{equation}
\begin{split}
\Psi (x,y,z,t) & = \psi(x,y,z) \exp{(-iEt\slash\hbar)} \sum_{n=-\infty}^{+\infty} J_{n} \left(\frac{eV_{\rm opt}}{\hbar\omega}\right) \exp(-in\omega t) \\
& = \psi(x,y,z)\sum_{n=-\infty}^{+\infty} J_{n} \left(\frac{eV_{\rm opt}}{\hbar\omega}\right) \exp -i(E + n\hbar\omega)t\slash\hbar\\
\end{split}
\label{eq:Psi} 
\end{equation}

where $J_{n}$ is the Bessel function of rank $n$. 

This new wave function suggests that an electron, initially at an energy $E$, can access a multitude of energy states separated by an energy ($\hbar\omega$). The wave function of the electron at energy $E + n$ is described by the Bessel function of order $n$ where $n$ corresponds to the number of photons absorbed or emitted by the electron in a multiphoton process.

The wave function in the presence of an optical field is then composed of energy terms $E$, $E \pm \hbar\omega$, $E\pm 2\hbar\omega$, etc. Each absorbed photon thus transfers its energy to an electron. Contrary to the classical approach where the variation of the current resulting from the AC disturbance evolves in a continuous way, the semi-classical approach induces an evolution of the $I(V)$ characteristic by $\pm\hbar\omega$ increments.
The optical field modulates the electron phase of one electrode with respect to the other, leading to a wave function with distinct energy states separated by the incident field energy $eV_{\rm opt} = \hbar \omega$. Thus, a bias $nV_{\rm opt}$ is added to the external bias $V$ at the rectenna terminal for each $n$ component of the wave function. The electronic density, relative to the squared wave function modulus $\Psi^{2}$, is then also proportional to the square of the Bessel function. Each component of the wave function is therefore weighted by a factor $J_{n}^{2} (\frac{eV_{\rm opt}}{\hbar\omega})$. 
The average current $I_{\rm rect}$ is then given by the relation describing the sum of all photonic absorptions:

\begin{equation}
I_{\rm rect}(V, V_{\rm opt}) = \sum_{-\infty}^{+\infty} J_{n}^{2} \left(\frac{eV_{\rm opt}}{\hbar\omega}\right) I_{\rm b} (V + nV_{\rm opt})
\label{eq:Psi2}
\end{equation}
$I_{\rm b}$ represents the tunnel current transported through the junction in the absence of excitation and is described by the equation~\cite{Bardeen1961}:

\begin{equation}
I_{\rm b} = C \int_{-\infty}^{+\infty} [f(E-eV) - f(E)] \rho_{1} (E- eV) \rho_{2} (E) dE
\end{equation}

where $C$ is a constant of proportionality, $V$ is the static bias applied to the nanojunction, $f$ the Fermi-Dirac distribution, $\rho_{1}$ and $\rho_{2}$ are the electronic states densities of each electrode. 
Thus, according to eq.~\ref{eq:Psi2}, the expression for the tunnelling current in the presence of an optical field becomes :
\begin{widetext}
\begin{equation}
 \langle I_{\rm rect} \rangle = C \sum_{-\infty}^{+\infty} J_{n}^{2} \left(\frac{eV_{\rm opt}}{\hbar\omega}\right)\int_{-\infty}^{+\infty} [f(E-eV) - f(E + n\hbar \omega)] \rho_{1} (E- eV) \rho_{2} (E + n \hbar \omega) dE   
\label{eq:Iopt}
\end{equation}
\end{widetext}

The argument of the Bessel function ($\frac{eV_{\rm opt}}{\hbar \omega}$) determines the occupation probability of the electron in the $n$-th electronic state. At low intensity and high frequency, the multi-photon absorption probability implies that only the terms corresponding to $n=-1, 0, 1$ present a significant contribution in the establishment of the photo-assisted tunneling current.
Consequently,
\begin{widetext}
\begin{equation}
\langle I_{\rm rect} \rangle = J_{0}^{2} \left(\frac{eV_{\rm opt}}{\hbar\omega}\right) I_{\rm b} (V) + J_{-1}^{2} \left(\frac{eV_{\rm opt}}{\hbar\omega}\right) I_{\rm b} \left(V - \frac{\hbar \omega}{e}\right) + J_{1}^{2} \left(\frac{eV_{\rm opt}}{\hbar\omega}\right) I_{\rm b} \left(V + \frac{\hbar \omega}{e}\right)
\label{eq:Iopt2}
\end{equation}
\end{widetext}

In the semi-classical regime, we can consider that $eV_{\rm opt} \leqslant \hbar \omega$. The argument of the Bessel function thus becomes much less than unity and the expression of the current $ \langle I_{\rm rect} \rangle$ can be simplified by carrying out a limited development of order 2 in the vicinity of 0. Setting $x = \frac{eV_{opt}}{\hbar \omega}$, then if $x ~\leqslant 1$,

\begin{equation}
\begin{split}
& J_{0} (x) \thicksim~ 1 - \frac{x^{2}}{4} + o(x^{4})\\
& J_{1} (x) \thicksim~ \frac{1}{2} x \left[ \frac{1}{1!} + \frac{-\frac{1}{4} x^{2}}{2!} \right] = \frac{x}{2} - \frac{x^{3}}{16} + o(x^{5})
\end{split}
\end{equation}

Moreover, 

\begin{align*}
J_{-n} (x) = (-1)^{n} J_{n} (x)
\end{align*}

Thus,
\begin{align*}
J_{-1} (x) = - \frac{x}{2} + \frac{x^{3}}{16} + o(x^{5})
\end{align*}

The equation \ref{eq:Iopt2} therefore becomes :

\begin{widetext}
\begin{equation}
\langle I_{\rm rect} \rangle = \left(1 - \frac{x^{2}}{4}\right)^{2} I_{\rm b} (V) + \left( \frac{x}{2} - \frac{x^{3}}{16}\right)^{2} I_{\rm b} \left(V + \frac{\hbar \omega}{e}\right) + \left(-\frac{x}{2} + \frac{x^{3}}{16}\right)^{2} I_{\rm b} \left(V - \frac{\hbar \omega}{e}\right) + o(x^{4})
\label{eq:DL}
\end{equation}
\end{widetext}

After removing terms higher than order 2, Eq.~\ref{eq:DL} simplifies and the tunnelling current appears as the sum of two contributions, the static tunnelling current and the photo-assisted tunnel current.

\begin{align}
\langle I_{\rm rect} \rangle & = \left(1- \frac{x^{2}}{2}\right) I_{\rm b} + \frac{x^{2}}{4} \left[I_{\rm b} \left(V + \frac{\hbar \omega}{e}\right) + I_{\rm b} \left(V - \frac{\hbar \omega}{e}\right)\right]\\
&= I_{\rm b} + \frac{1}{4} V_{\rm opt}^{2} \left[\frac{I_{\rm b}\left(V + \frac{\hbar \omega}{e}\right) - 2I_{\rm b}(V) + I_{\rm b}\left(V- \frac{\hbar \omega}{e}\right)}{\left(\frac{\hbar \omega}{e}\right)^{2}}\right]
\label{eq:semiclassique}
\end{align}

In the low frequency limit, Eq;~\ref{eq:semiclassique} leads to the classical expression of the rectified current (Eq.~\ref{Eq:Irect}). The quantized steps are observable when $\hbar \omega/e$ is of the same range of magnitude or greater than the bias and corresponds to a high nonlinearity of $I(V)$. The classical description is therefore an extension of the quantum study of the rectification phenomenon.

%\nocite{*}

\end{document}